\documentstyle[11pt]{article}

\begin{document}

\title{A response to the commentaries on CoRR%
\thanks{This work was supported in part by the NSF, under grant
IRI-96-25901.}}
\author{Joseph Y. Halpern\\
Computer Science Department\\
Cornell University\\
halpern@.cornell.edu}
\date{\today}
\maketitle

I thank Les Carr, Wendy Hall, Steve Hitchcock, Stevan Harnad,
David Armbruster,  James Prekeges, and A.~J. van Loon for their
comments on my article.  I agree with most of the points made
by Carr et al.,
Armbruster, and Prekeges; while I think van Loon makes some
interesting points, my impression is that he has a number of serious
misconceptions of how CoRR works and  some deep
misunderstandings of what the research community wants and what matters
to us (at least, in the
fields that I am most familiar with---computer science, mathematics,
physics, and economics).  Let me respond to their
comments in turn.

First, with regard to Carr et al., it goes without saying that I
strongly support the Open Archives initiative.  Turning to their
comments on CoRR itself, I must confess that their figure of roughly
1600 papers on CoRR as of December, 1999, is correct.  (It seems that I
simply added incorrectly; so much for having a Ph.D. in mathematics
\ldots) I think that the comparison with NCSTRL's 27,000$+$ is a bit unfair---NCSTRL
has papers dating back to 1958.  However, this number is useful insofar
as giving an idea of how many papers CoRR could potentially have.
With regard to their more substantive comments regarding why
more authors
have not submitted (yet) to CoRR, I agree strongly with their first
suggestion, that we need more effective promotion.  We are planning
another round of promotion as soon as a slightly improved user interface
is installed (this is awaiting the return of some of the LANL staff from
vacation, and may well have happened by the time this article is
published).  I would certainly welcome suggestions for how to do more
effective promotion.

Their second suggestion is that we get
stronger support from our sponsors.  While it is true that ACM has been
somewhat ambivalent in its support---it is, perhaps understandably,
concerned about the impact of CoRR on its journal publications---it is
not clear exactly what kind of support
Carr and his colleagues have in mind.  I can think of
a few things that ACM could (and, I believe, should) do, such as
building a better gateway to CoRR from the
ACM web site, promoting CoRR on the ACM web site,
and encouraging editors-in-chief of their journals to
encourage authors to submit papers by posting them on CoRR (something
which I am doing as editor-in-chief of the {\em Journal of the ACM\/}
and is also being done for the new {\em ACM Transactions on Computational
Logic}).  Similar requests could be made of other sponsors.
I would certainly
welcome other concrete suggestions.  It is much easier to ask for
something specific than to ask for vague expressions of stronger support.

As far as having
a clearer relationship with journals, as I said in my original article,
I have been in contact with most of the major publishers in computer
science.  While all I have checked with allow authors to post preprints
on CoRR, none seem interested in working with CoRR, with the exception
(with some ambivalence, as noted above) of ACM.  There is a smaller
journal, the {\em Journal of AI Research}, that has agreed to post all
its papers on CoRR.  While I am pursuing this further, I also suspect
that the impact of journal cooperation will not be significant.

My own
strong belief is that the problem is not lack of cooperation or with
publishers or sponsors, but with lack of awareness, author
inertia, and user interface.  As a result, I am focusing on improving
the interface, promotion, and on increasing the size of CoRR by
incorporating
some pre-existing archives (one from AT\&T with 60 papers will shortly
be incorporated; it will take a little longer, but within the year I
also hope to include over 700 papers on Software Engineering from the
Software Engineering Institute at Carnegie Mellon) and
 encouraging the use
of CoRR to house online conference
proceedings.  The Workshop on Nonmonotonic Reasoning is in fact putting
its proceedings on CoRR.
As a result, there were 66 submissions to CoRR in the first 15 days of
March alone, which is 50\% more submissions than in any other month
since the first month that CoRR opened.  My hope is that this approach
will lead to a momentum effect.
Once people start posting papers on CoRR, others will follow suit.

Of course, it could be, as Carr et al.~suggest, that I and my colleagues
on the CoRR committee simply do not understand our community well.
All I can say is that between the CoRR committee and the subject area
moderators, we have almost 50 leading computer scientists.  I believe
that this is enough of a cross-section to at least give us a reasonably
good understanding of what authors in computer science want.
Having
said that, though, I certainly welcome the suggestions made by Carr et
al., and would be interested in hearing others.

Turning to Armbruster's comments, I agree that
there is no question that different cultures will respond to
online publishing in different ways.  I hope that Armbruster's editorial
aside---that publishers will figure out a way to continue making money
while authors submit their research findings to online repositories---is
true.  It is certainly the case that publishers (including ACM)
are scrambling to find such a way, although I don't believe that any
have found it yet.

Armbruster also raises the issue of what papers should be included in
the repository (drafts? preprints? peer-reviewed manuscripts?).
My short answer to this is ``all of them.''   But this short answer
hides some potential complexity.  Let me explain.
A paper evolves through a  number of stages; the exact stages are
somewhat field dependent.  In computer science, the typical stages are
(rough) draft, preprint (or, more likely these days, eprint),
conference publication, and journal publication.  There may well be
several versions of the preprint.  Authors must decide at
which point(s) the paper should be ``checkpointed.''  Is it ready to
bring out as a preprint?  Is it worth bringing out a new version of the
preprint?  Is it ready to submit for journal publication?

CoRR was intended to focus mainly on the preprint stage, but
authors can certainly post a paper in any stage of development.
What is to stop authors from cluttering up the archive with very early
and incomplete versions of a document?  Nothing, just as there is
nothing stopping an author from bringing out a technical report at any
stage in a paper's development and then bringing out frequent revisions.

Of course, in practice, authors typically don't publish very early
drafts of a paper nor do they revise technical reports all that
frequently.   The reasons are easy to understand: early drafts are felt
to be too premature to be made available to a wide readership and
frequent revisions will be ignored by already overloaded readers.
Interestingly, the rule that authors have only 24 hours to withdraw a
paper, far from encouraging the submission of immature material, as
suggested by van Loon, is intended to prevent the submission of immature
material, since once submitted, an early draft remains on CoRR as an
embarrassing reminder.  Anecdotal evidence suggests that the rule does
indeed have this effect.
Experience with the physics archive at LANL (for which there is much
more data than for CoRR), suggests that authors indeed post relatively
few versions of their paper, and post papers at all stages in their
evolution.

This discussion so far has implicitly assumed that the evolution of a
paper is essentially linear.  Later versions subsume earlier
versions.  For such papers, the CoRR mechanism works well.
All versions are timestamped with the time of their submission, making
it easy to decide which is later (even if they are published in the same
year---the concern by van Loon simply regarding the ``definitive''
version simply disappears in the online world, if the latest version is
always taken to subsume earlier ones).   However, later versions may
{\em not\/} always subsume earlier ones.  Indeed, take the case of this
paper.  It has appeared in three versions.  The earliest one (written at
the request of William Arms, editor-in-chief of {\em D-Lib Magazine\/},
to promote CoRR when it first started) is indeed subsumed
by the later two, but the later two are incomparable.  The version that
appears in the {\em Proceedings of the 1999 ACM Digital Libraries '99\/}
contains some technical discussion of how interoperability was achieved
between NCSTRL and LANL through the Dienst protocol, which I judged
would be of less interest to the {\em SIGDOC\/} community.  On the other
hand, the version of the paper in this issue contains some discussion
of issues such as preservation and participation that were written in
response to comments by Dr.~Girill, the editor-in-chief of {\em SIGDOC}.

While I could imagine a version of this paper that subsumed
all the currently-existing versions, in general this may not always be
possible or even desirable.  For example,
there are some papers of mine that I would like to
target to both economists and computer scientists.  However, the points
that I would emphasize to economists are quite different from those I
would emphasize to computer scientists; in addition, the background that
I can assume from one community would be very different from the other.
The current publication structure does not provide a convenient solution
to this problem.  It is considered unethical to submit the same or
very similar papers to two different journals, so I must essentially
choose which community to address the paper to.  In the online world,
a number of solutions are available which are not as easily available
in the paper world.  By using HTML, it is possible to write a document
in such a way that each of two communities can click on the material
appropriate for them.  Alternatively, it seems quite reasonable
to have different versions of a paper with a common core targeted for
different communities.  The commonality could be made quite apparent
online in ways that cannot be done on paper.  While CoRR is not yet set
up to deal in an optimal way with this issue, I see no intrinsic
difficulties in doing so.

James Prekeges raises a very important issue that I think can be best
thought of, not in terms of censorship, but {\em filtering}.  CoRR does
only minimal filtering. As I said in my article, we just make sure that
papers are relevant to the subject area in which they are submitted.
While this could act as censorship, in practice it has not.  (Papers can
always be submitted to the ``Other'' subject area, which is explicitly
intended for papers that do not fit anywhere else.)  Since readers can
subscribe to subject areas (which means that they get email notification
of any new papers that are posted in that subject area), the intent of
this minimal filter is just to ensure that readers are not notified of
too many papers that they view as irrelevant.  Prekeges is certainly
right that readers will want more of a filter.

One obvious filter is peer review.
The fact that a paper has been certified by an editorial
board (or some other certifying authority) can easily be noted on CoRR.
Indeed, the online structure makes it reasonable for a paper to be
certified by various certification boards.  For example, if I write a
multidisciplinary paper that is intended for both computer scientists
and economists, it seems to me perfectly reasonable to ask both
communities to certify it: the computer science
community for its computer science content and the economics community
for its economics content.%
\footnote{It seems less reasonable to me to have the paper certified by
two different boards in computer science.  This is a waste of
reviewers.}

But let us be clear that peer review is just one form---and a rather
imperfect one at that (although perhaps the best we have now)---of
{\em certification}.  By accepting a paper
to the {\em Journal of the ACM}, I am
certifying that it meets {\em JACM}'s rather stringent standards of
quality control.  But that judgment is largely based on the reviews of
two reviewers.  Prekeges mentions another form of filtering: a
users' feedback facility such as that provided by amazon.com.  As I
mentioned in my original article, it would certainly be
possible to build a comment facility on top of CoRR, and I suspect it
will be done at some point.  However, experience with other attempts to
do just that has shown that (at least so far) such comment facilities
have been used relatively little.  Perhaps the {\em Digital Review\/}
experiment that I mentioned in my original article will be more
successful.

It is also possible to certify authors in various ways.  By clicking on
the name of an author of a paper on CoRR, it is already possible to
see what other papers the author has submitted to CoRR.  If CoRR becomes
the standard place for authors to submit, then this will give readers
some idea of an authors' profile of publications.
Services such as the {\em Science Citation Index\/}
also certify authors, by showing how often their papers have been
referenced.  There is a free online analogue by Bollacker, Giles, and
Lawrence called citeseer, available at http://citeseer.nec.nj.com, which
I am hoping will shortly be integrated with CoRR.  In summary, there
already is some minimal filtering information available on CoRR, and I
expect that more and more will be available, probably sooner rather than
later.  The issues raised by all the commentators on this score, while
legitimate, will not, I believe, cause problems in the long (or even
short) run.

Finally, let me respond to some of the issues raised by van Loon not
already addressed above.
\begin{itemize}
\item I believe that
van Loon is confounding two issues when he speaks of {\em publication}.
I will reserve the word ``publication'' for its original meaning:
``making public''.  Another important aspect of journal publication, as
suggested above, is {\em certification}.  Authors both want to
make their paper publicly available (one hopes that we actually do
research in order to influence others, after all) and to get it
certified (for tenure and promotion, grant proposals, and so on).  CoRR
is intended to facilitate rapid publication.  In a field moving as
rapidly
as computer science, authors understandably want to get their ideas out
quickly, to get feedback.  As I suggested above and in my original
article, CoRR can also be used to facilitate certification.  But it is
important to decouple these two objectives.
\item In Section 2.1 of his commentary, van Loon suggests that the
reason to publish electronically is
``cost saving in the long term''.  While this is certainly a factor, it
is not the one that motivates most of the research community.
Electronic publication (again, I stress that by ``publication'' I mean
``making public'', not necessarily certification) makes papers
quickly and easily accessible in ways that journal publication does not.
Although the Cornell library subscribes to many journals, it does not
subscribe to all of them and, in any case, many of the articles I am
most interested in are not (yet) in the journal literature.  And even if
they are, it is much more convenient for me to get an article on the
web than to get it from a journal in the library.
\item In Section 2.2 of his commentary, van Loon suggests that ``This
option [of having unrefereed papers on CoRR] is acceptable for
researchers \ldots only if CoRR is not meant to be reliable, but only
informative of what people are doing.''  Perhaps this is the case in some
fields, but it is certainly not how things proceed in computer
science and other fields with which I am familiar.
Many of the papers that I read and refer to in my papers are unpublished
at the time I read them.  In a field moving as rapidly as computer
science, I cannot wait until they have been certified to read them.  So
how do I know which papers to read?  The obvious ways: reputation of
author, recommendations from colleagues, references in other papers, a
quick scan of the abstract and introduction for relevance.   In my areas
of specialty, I am usually quite capable of homing in on what I need
quite quickly.  Even outside my area, with the help of colleagues, I can
usually find what I need.  I am not at all unusual in this regard.
\item In Section~3.1, van Loon quotes me accurately as saying that we rejected
the option of joining LANL because ``it did not provide an interface to
which other repositories could join''.  He then goes on to say ``One
must assume that Halpern not only established this fact, but has also
tried to convince Los Alamos people to provide a suitable interface.
This attempt was apparently in vain.''  Nothing could be further from
the truth!  As perhaps I didn't make clear enough in my article,
CoRR is actually part of the LANL archive.  The url for CoRR is in fact
http://xxx.lanl.gov/archive/cs.  As a result of a sequence of meetings
between people from LANL and NCSTRL, the LANL software was modified to
be compatible with the Dienst protocol used by
NCSTRL to provide an open architecture.  (This is discussed in more
detail in my paper with Lagoze in {\em ACM Digital Libraries '99}.)
Consequently, all the computer science material submitted to LANL is in
fact on CoRR.  Moreover, the use of the Dienst protocol allows us to
link NCSTRL and CoRR (and, as I mentioned in my original article, allows
CoRR to be a node on NCSTRL).  In fact, as was mentioned by Carr et
al., the use of Dienst is
critical in the plan to build a federation of online repositories (see
http://www.openarchives.org and H. Van de Sompel and C. Lagoze, ``The
Sante Fe Convention of the Open Archives Initiative'', {\em D-Lib
Magazine}, Feb., 2000, available at
http://www.dlib.org/february00/02contents.html).  This federation
should lead precisely to the
interdisciplinary superdatabase envisaged by van Loon in Section 3.2.
The database will contain not only pointers to where the data can be
retrieved,
but the actual documents.  I view this as perhaps the most exciting
development in the area in the past few years.
\item In Section 4, van Loon states concerns about the long-term
stability of CoRR for financial reasons, and suggests imposing a
downloading charge.   The implication is that this should be done right
away.   It is interesting that, while van Loon suggests the CoRR
project as a whole is premature and needs further discussion, he does
not seem to think that the idea of imposing downloading charges is
premature.  While it is certainly conceivable that imposing a
downloading charge will be necessary for long-term stability, I doubt
that this is the best solution.
My own feeling here is that the most important attribute for
ensuring the longterm stability of CoRR is making sure that it houses a
lot of documents.   If it does, then it will be a sufficiently important
resource for the community that a way will be found to ensure its
survival.  I should add that, in my opinion, the long-term
stability of for-profit publishers is in at least as much doubt as that
of CoRR (given the anticipated shakeout likely to be caused by web
publication).  Some will no doubt find a way to survive in the brave new
world; others will not.
\end{itemize}

As I indicated in my original article,
the CoRR project was discussed for over a year by leaders in
the computer science community.  I certainly have no qualms about making
it available when we did.  However,
let me conclude by agreeing with one observation made by van Loon.  I
submitted my article to {\em SIGDOC\/} in part because I did not have
enough confidence that it would reach the right audience if I just
posted it on CoRR.  I hope, of course, that will change soon.  In
the mean time, I do plan to post the article and the response on CoRR.
(As I said in my original article, this is allowed by ACM's copyright
policy.)  I encourage the other commentators---and all other members of
the {\em SIGDOC\/} community!---to do the same.
\end{document}